\documentclass[a4paper, fleqn, usenatbib, useAMS]{mnras}

\def\Ms{M_{\rm sat}}
\def\Mh{M_{\rm host}}
\def\Mo{M_\odot}
\def\mr{\Ms/\Mh}

\def\aap{AA}
\def\apjl{ApJL}
\def\apjs{ApJS}
\def\mnras{MNRAS}
\def\apj{ApJ}
\def\aj{AJ}

\def\nat{Nat}
\def\araa{ARAA}
\usepackage{afterpage}
\usepackage{multicol}        
\usepackage{graphicx}
\usepackage{float}
\usepackage{bm} 
\usepackage{amssymb}
\usepackage{hyperref}
\usepackage{amsmath} 
\usepackage{subfig} 
\usepackage{pdflscape}
\usepackage[export]{adjustbox}
\usepackage{mathrsfs} 
\usepackage{mathtools}

\title[GC populations \& minor mergers]{Globular cluster populations and the kinematical fingerprints of minor mergers}
\author[N. C. Amorisco]{N. C. Amorisco$^{1, 2}$\thanks{E-mail:
nicola.amorisco@cfa.harvard.edu} \\
$^{1}$Institute for Theory and Computation,  Harvard-Smithsonian Center for Astrophysics,  60 Garden St.,  MS-51,  Cambridge,  MA 02138,  USA\\
$^{2}$Max Planck Institute for Astrophysics,  Karl-Schwarzschild-Strasse 1, 85748 Garching, Germany}

\begin{document}



\maketitle

\label{firstpage}

\begin{abstract}
We use Monte Carlo $\Lambda$CDM assembly histories, minor-merger N-body simulations and empirical
relations between halo mass and the globular cluster (GC) abundance to study the kinematical properties of 
halo GCs in massive galaxies, with $M_{\rm vir}(z=0)=10^{13.5}\Mo$. 
{While the accreted stellar halo is dominated by the contributions of massive satellites, we show that
satellites with low virial mass (i.e. low satellite-to-host virial mass ratio, VMRs) are important contributors to the population of accreted GCs. The relative contribution of accretion events with low VMRs is highest for the halo population of blue GCs and gradually decreases for red GCs and accreted stars.}
As a consequence of the reduced efficiency of dynamical friction on minor mergers, our populations of 
cosmologically accreted blue GCs are systematically more {spatially extended} and have 
higher velocity dispersions than for red GCs, in agreement with observations. 
For the same reason, assembly histories including a higher fraction of minor mergers result in more {spatially extended} GC populations. 
GC line-of-sight velocity distributions featuring negative values of the kurtosis $\kappa$, as recently observed, 
are ubiquitous in our models. Therefore, $\kappa<0$ is not at odds with an accretion scenario, 
and in fact a fingerprint of the important contribution of minor mergers. However, our populations of accreted GCs 
remain mostly radially biased, with profiles of the anisotropy parameter $\beta$ that are 
mildly radial in the center ($\beta(r<10~{\rm kpc})\sim0.2$) and strongly radially anisotropic at large 
galactocentric distances ($\beta(r>30~{\rm kpc})\gtrsim0.6$), {for both red and blue populations}.  

\end{abstract}

\section{Introduction}

Globular clusters (GCs) are valuable tracers of the kinematics and assembly of galaxies. 
Compact and bright, they are observable in distant galaxies {in the local universe} as well as at large radii from the galaxy centre.
As such, GCs have had a fundamental role in studies aimed at probing the dark matter distribution in massive early type 
galaxies \citep[e.g.,][and references therein]{Gr94, RS00, RK01,Sch12, AA14, NN14, VP15, 
Zhu16} or at constraining their assembly history \citep[e.g.,][among many others]{Ze93,DF97,PC98,JB06,LG14,PM16}.

Intriguingly, GC populations appear to have strong ties to dark matter haloes. 
{It is well known that the efficiency of the process of galaxy formation is a 
markedly non-monotonic function of halo mass,
with a characteristic peak at around Milky Way mass haloes \citep[e.g.,][]{Guo10,BM10,PB13}.}
In turn, the GC abundance -- or 
the total stellar mass in GCs -- appears to be roughly proportional to halo mass
\citep[e.g.,][]{Bl97, LS09, IG10, Hu14, Ha17}. Such an approximately linear relation seems to extend
to low-mass haloes \citep{DF18}, so that, within a $\Lambda$CDM framework, it directly follows
that a considerable fraction of GCs in massive galaxies have not been formed 
within the main progenitor, but rather they have been accumulated through hierarchical 
merging \citep[e.g.,][]{WF91}. In fact, a close to linear relation between halo mass and GC abundance works well with 
the scenario in which most GCs, and especially metal poor GCs, form at high redshift 
\citep[e.g.,][and references therein]{Pee84,MR02,KG05,JB06}, as hierarchical 
assembly would approximately preserve this linearity.

The kinematics of accreted stellar populations deposited in the halo of galaxies through
hierarchical accretion has been studied in detail, with particular attention to the case of Milky
Way (MW) mass galaxies \citep[e.g.,][]{AD05,MA06,BM06,On07,AC10,Wu14}. A general result of these analyses is that
radial orbits are dominant, and that the deposited stellar material is therefore characterized by a 
marked radial anisotropy, $\beta>0$, where $\beta(r)$ is the classical anisotropy
parameter
\begin{equation}
\beta(r)=1-{{\sigma^2_t(r)}\over{2 \sigma^2_r(r)}} \ ,
\end{equation} 
with $\sigma_t$ and $\sigma_r$ being respectively the tangential and radial 
stellar velocity dispersions. Given the generality of this finding and the common accretion origin of halo GCs,
it would seem only natural that the population of accreted GCs should share this property, and also have radially biased orbits. 

The recent increase in both size and quality of GC spectroscopic datasets \citep[e.g.,][]{VP13,ZP15,DF17}
have however sparked a question on whether the simple picture above may be incomplete. For instance, 
following the described line of though, the line-of-sight velocity distribution (LOSVD) of accreted GCs should be characterised by a positive kurtosis, as generally expected for populations with orbital distributions that have a strong
radial bias \citep[e.g.,][]{OG93,vdM93}. {For a LOSVD $f_{LOS}(v,R)$, the kurtosis $\kappa$ 
quantifies departures from Gaussianity that are symmetric with respect to the mean $\mu$:
\begin{equation}
\kappa(R)+3={1\over {\sigma_{\rm LOS}^4(R)}} \int dv {f_{LOS}}(v,R) \left[ v-\mu(R)\right]^4 \ .
\end{equation} 
Here $R$ is the projected galactocentric radius, $\sigma_{\rm LOS}(R)$ is the LOS velocity dispersion, 
and the LOSVD $f_{LOS}(v, R)$ is indeed the distribution of LOS velocities at the radius $R$, normalised to unity}.
Radially biased populations are characterised by LOSVDs with heavy tails and a sharp central peak, 
corresponding to $\kappa>0$ \citep[see e.g.,][]{NA12}.
Instead, \citet{VP13} have shown that both red and blue GC populations
often display LOSVDs with kurtosis $\kappa\sim0$, suggesting orbits that are close to isotropic.
Positive values of $\kappa$ are often observed at large galactocentric radii,
but negative values of $\kappa$ are common for both red and blue
GCs in the galaxy's central regions, as also confirmed by \citet{ZP15}. 
{This is unexpected and, if taken at face value, this finding could be interpreted as a suggestion 
of some degree of {\it tangential} anisotropy. Such a conclusion would be unwarranted: it is well known that 
kurtosis is only a rough proxy for actual orbital structure, and that there's no one-to-one relation between kurtosis 
and anisotropy \citep[e.g.,][]{OG93,vdM93,NN14}. However, the finding that negative values of $\kappa$ are common in GC populations remains surprising
when put in comparison with the stellar halo.} Generalizing from the kinematic properties
of the latter, negative values of $\kappa$ would appear at odds with the scenario in which a substantial 
fraction of all GCs are accreted. 

\begin{figure*}
\centering
\includegraphics[width=\textwidth]{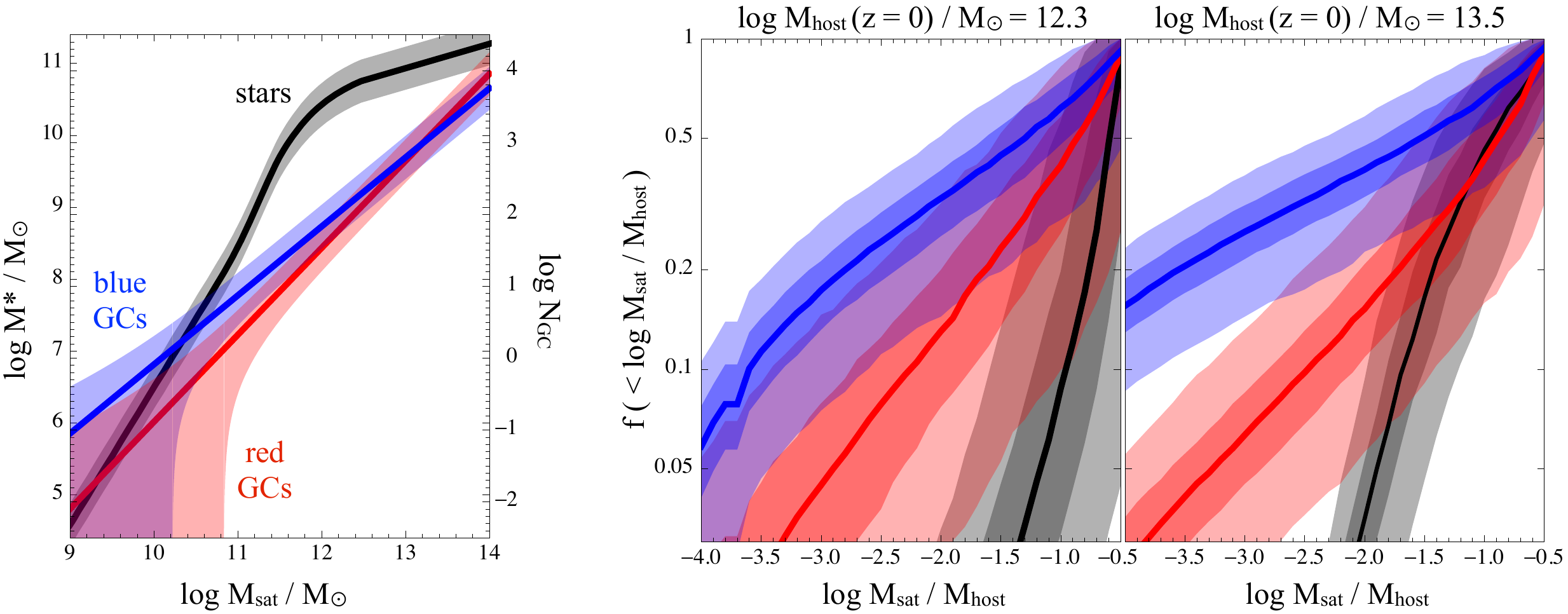}
\caption{Left panel: the relation between halo mass of the contributing satellites and 
the contributed stellar mass (in black), red GCs (in red), blue GCs (in blue). The shading 
identifies the 1-sigma regions around the mean model relations (including both intrinsic 
and Poisson scatter for GCs). Middle and right panels: cumulative contributions to the 
accreted stellar and GC populations, as a function of the virial mass ratio at accretion, 
$\log \mr$. Thick lines identify medians, shaded regions extend between the $\{5, 25, 75, 95\} \%$ quantiles,
over a sample of $10^3$ halo assembly histories. The middle panel refers to the halo of a
Milky Way-like galaxy $\log\Mh(z=0)/\Mo=12.3$. The right panel to a massive elliptical galaxy 
$\log \Mh(z=0)/\Mo=13.5$.}
\label{fig1}
\end{figure*}

In this paper, we show that there is in fact no contradiction between the observed negative values of the
kurtosis in the LOSVDs of halo GCs and the scenario in which most of them have an accretion origin. We
show that negative values of the kurtosis are expected within a $\Lambda$CDM framework when minor 
mergers are the dominant contributing channel of an accreted population, without the need to invoke 
additional physical processes. Due to the approximate linearity of the relation 
between halo mass and GC abundance, and in opposition to stellar haloes, minor mergers are indeed dominant in assembling the accreted 
populations of GCs, especially for blue metal poor GCs. 

In Section~2 we use $\Lambda$CDM assembly histories 
to show how stellar and GC haloes have systematically different mean progenitors. 
Section~3 concentrates on the kinematical properties of material deposited in the 
accreted halo by satellites with different properties. 
Section~4 builds toy models for the populations of cosmologically accreted GCs, 
and focuses on their intrinsic and projected kinematic properties. 
Section~5 discusses results and lays out the Conclusions.

\section{The progenitors of stellar and GC haloes}

{We construct two sets of mock $\Lambda$CDM assembly histories, one for haloes of 
MW-like virial mass, $\log\Mh(z=0)/\Mo=12.3$, and one for haloes with virial mass $\log \Mh(z=0)/\Mo=13.5$,
representative of a massive central galaxy. Each set explores $10^3$ individual halo
assembly histories. Individual assembly histories are composed of Monte Carlo-generated accretion events,
randomly sampling from the merger rates of dark matter haloes in the Millennium simulation \citep[]{Fak10}, 
and collect all mergers experienced by the main progenitor. Statistically, these assembly histories reproduce 
the growth of haloes in the Millennium simulation, in terms \citep{Fak10} of the distribution of accretion redshifts
and of satellite-to-host virial mass ratios, in both mean and scatter.}
We keep track of all accretion events onto the main progenitor for satellites with virial mass $\log \Ms/\Mo>7$ at the time of accretion, since $z=6$. 
These accreted satellites contribute stars to the accreted halo population of the host according to a standard 
stellar-to-halo mass relation (SHMR). In particular, we adopt a mean relation as presented by \citet{GK14},
and assume a log-normal scatter of $0.3$~dex {(individual values are randomly sampled from such
distribution)}. For simplicity, both mean relation and scatter are assumed 
to be independent of redshift. The resulting SHMR is displayed as a black line in the left panel of Fig.~1. 
{We recall that below the break at MW-like masses, the log-log slope of this SHMR is of $\approx 1.9$}.

Accreted satellites also contribute to the accreted GC population of the host. 
We track red and blue GCs separately and assume that the GC population of each satellite 
is a function of the halo mass at accretion. In particular, we adopt mean relations\footnote{{
We recall that these relations refer to the integrated GC population. The measured 
GC abundances used for their derivation are therefore based on a number of hypotheses regarding the 
GC luminosity function, as well as on the spatial density distribution of the GC population when the field 
coverage does not include the entire target galaxy. We refer the reader to \citet{Ha13} and \citet{Jo07} for more details.}} as measured 
by \citet{Ha15}:
\begin{equation}
\left\{\begin{array}{lll}
\log N_{GC,red}&=& 1.8 + 1.21\left(\log \Ms-12.2\right)\\
\log N_{GC,blue}&=& 2.0 + 0.96\left(\log \Ms-12.2\right)
\end{array} \right. \ ,
\label{NGCrel}
\end{equation} 
{where $N_{GC,red}$ and $N_{GC,blue}$ are respectively the abundance of red and blue GCs.}
With a log-log slope of $0.96$, the abundance of Blue GCs is approximately proportional to the 
virial mass of the satellite $M_{sat}$. The abundance in red GCs is slightly steeper, for a log-log slope of $1.21$. 
We take that, in addition to Poisson noise, the intrinsic scatter around these mean relations is of $0.3$~dex.
The left panel of Fig.~1 compares the prescriptions above with the SHMR; the shading shows the 1-sigma region for each relation. 
It is worth noticing that while the empirical relations~(\ref{NGCrel}) have been measured at $z~0$, 
we instead require estimates of the GC abundance of satellites at the time of accretion. 
In absence of measurements for the same relations as a function of redshift, we take 
that the abundances~(\ref{NGCrel}) are independent of redshift: satellites contribute GCs according to their 
halo mass at accretion. Note that this is the same simplifying assumption made for the SHMR. 
Though not ideal, this can be considered conservative here. With respect to a model in which GCs are formed 
at high redshift with an abundance that is proportional to halo mass {\it at the time of formation} 
\citep[see e.g.,][]{BK17}, the present model implies an underestimation of the role of 
minor mergers, as massive haloes assemble more recently. 

Given the set of minimal hypotheses above, Figure~1 shows the cumulative fraction $f$
of accreted halo populations, in stars and GCs, contributed by satellites with different
satellite-to-host virial mass ratio at accretion (VMR), $\log \mr$. The middle panel refers to 
our set of MW-like haloes. The right panel to our massive haloes with $M_{\rm vir}(z=0)=10^{13.5}\Mo$. 
Thick lines show median relations, the shaded regions extend between the $\{5, 25, 75, 95\} \%$ 
quantiles. Fig.~1 shows very clearly that the average progenitor satellite of stellar halo
and accreted GC populations are remarkably different.

\subsection{MW-like haloes}

As shown in the middle panel of Fig.~1, the contribution of low-mass satellites to the accreted stellar halo of MW like galaxies drops
quickly at $\log \mr\lesssim-1$. This is a well-known result \citep[e.g.,][]{BJ05,LS07,AC10,AD16,NA17} 
and a direct consequence of the steepness of the SHMR at virial masses $\log\Ms/\Mo\lesssim12$, 
with a log-log slope of $\sim1.9$ \citep[e.g.,][]{GK14,GK17,PJ18}.
As shown by the first panel of Fig.~1, the relations~(\ref{NGCrel}) prescribe GC abundances 
that are considerably less steep with halo mass, resulting in much higher fractions
of accreted GCs being contributed by satellites with low VMR. For example,
accretion events with $\log \mr<1/50$ contribute between 30 and 48\% (respectively 25 and 75\% quantiles)
of the blue GCs in MW-like galaxies {respectively between 10 and 27\% of the red GCs}. 
This is in stark contrast with the $<$3\% contributed to the stellar 
halo by the same satellites (75\% quantile). 

It is interesting to note that, in MW mass galaxies, the median distribution of VMRs of 
satellites contributing to the stellar halo, to the accreted population of 
red GCs and to the accreted population of blue GCs are all systematically different  from each other:
{the accreted GC population does not really trace the assembly of the stellar halo}.
As already mentioned, the stellar halo is dominated by high VMR accretion events. 
Minor mergers become increasingly important for red and blue GCs. As we will show in the following, 
this implies that the spatial distribution and kinematic properties of these three halo populations 
should be expected to be different. 

In this simple model, haloes with $\log\Mh(z=0)/\Mo=12.3$ accrete a total of between 149
and 196 blue GCs, and 31 to 49 red GCs (25 and 75\% quantiles for both populations).
Given the simplistic hypotheses on which they are based, these figures should  
be taken as qualitative estimates. However, it is interesting to notice that the simplest
possible combination of physical ingredients, i.e. $\Lambda$CDM assembly histories 
and the measured GC abundances~(\ref{NGCrel}) appear to produce $z=0$ GC populations that 
compare favorably with expectations. According to equations~(\ref{NGCrel}), a virial mass of
$\log\Mh(z=0)/\Mo=12.3$ corresponds to 83 red and 125 blue GCs (a scatter of 0.3~dex corresponds to 
approximately a factor of 2). Both total GC abundance \citep[see e.g.,][]{Ha17} and ratio between 
red and blue GC populations \citep[][]{Ha15} are roughly reproduced.

\subsection{Massive haloes}

The case of haloes with virial mass $\log \Mh(z=0)/\Mo=13.5$ is illustrated in the rightmost 
panel of Fig.~1. The relation between the abundance of blue GCs and halo mass
is the least steep, so the fractional contribution of minor mergers is highest for blue GCs,
as for the case of MW mass haloes. Between 39 and 55\% of blue GCs (25 and 75\% quantiles)
are contributed by satellites with $\log \mr<1/50$. 

Fig.~1 shows that the contribution of low-mass satellites to the accreted stellar halo 
of high-mass galaxies is also higher than in the case of MW-like haloes. This is a consequence of the break
in the SHMR at $\log \Mh(z=0)/\Mo\sim12$, and a well known result \citep[e.g.,][]{Pu07,TN07,AC13,AP14}.
{Differently from the case of MW-mass galaxies, in which all populations have different median 
trends, here the median cumulative fraction of accreted material in stars and red GCs 
are very similar to each other for mass ratios $\log \mr\gtrsim -1.2$. This corresponds to 
65\% of the contributions to these populations. In other words, in massive galaxies, a majority 
of the material in the accreted stellar halo and in the accreted population of red GCs often have 
similar progenitors.} This provides a qualitative justification to the finding that red GCs often trace 
the distribution of stars not just in the center, but also out to large galactocentric distances in massive galaxies \citep[e.g., ][]{JS11,VP13, AA14}. 
At these radii this behaviour can not be simply due to material formed {\it in situ},
and has to be related to the similar kinematical properties of accreted stars and red GCs. Note however, that the scatter 
in the distributions of VMRs is significant for both these populations, suggesting that this similarity may be loose
in some systems. For completeness, the total number of accreted blue GCs estimated by 
this simple model is of between 2350 and 2950, and of between 850 and 1250 for red GCs 
(25 and 75\% for both populations), again in rough agreement with the expectation for this 
halo mass (respectively 1770 blue GCs and 2360 red GCs). 

Figure~2 concentrates on the size of the different contributions to the accreted GC populations: 
$N_{GC,{\rm sat}}$ on the $x$ axis is the number of GCs contributed by an individual satellite. 
The top panel shows a demography of all contributions in terms of the number of accreted satellites
that contribute $N_{GC,{\rm sat}}$ GCs, since $z=6$. Points indicate medians, while the error bars extend
between 10 and 90\% quantiles of the distribution. Satellites that contribute $>$100 GCs are rare,
for both blue and red GCs. Due to preponderance of low mass accretion events, in these simple models, there 
are on average more satellites contributing a single GC then satellites contributing $N_{GC,{\rm sat}}>1$.
This is mirrored in the lower panel, which shows the total number of GCs accreted in a contribution 
of size $N_{GC,{\rm sat}}$. {Note that, if the upper panel plots the function $N_{sat}(N_{GC,sat})$,
the lower panel is plotting $N_{GC}=N_{GC,sat}\times N_{sat}(N_{GC,sat})$}. Poisson noise in the assembly 
histories is especially
evident: high VMR mergers (e.g., $\mr>0.5$) are rare, but may contribute a large
fraction of all GCs, introducing significant stochasticity. The number of GCs contributed 
individually or as part of small groups shows less scatter, and is consistently high in all assembly histories. 

\begin{figure}
\centering
\includegraphics[width=.8\columnwidth]{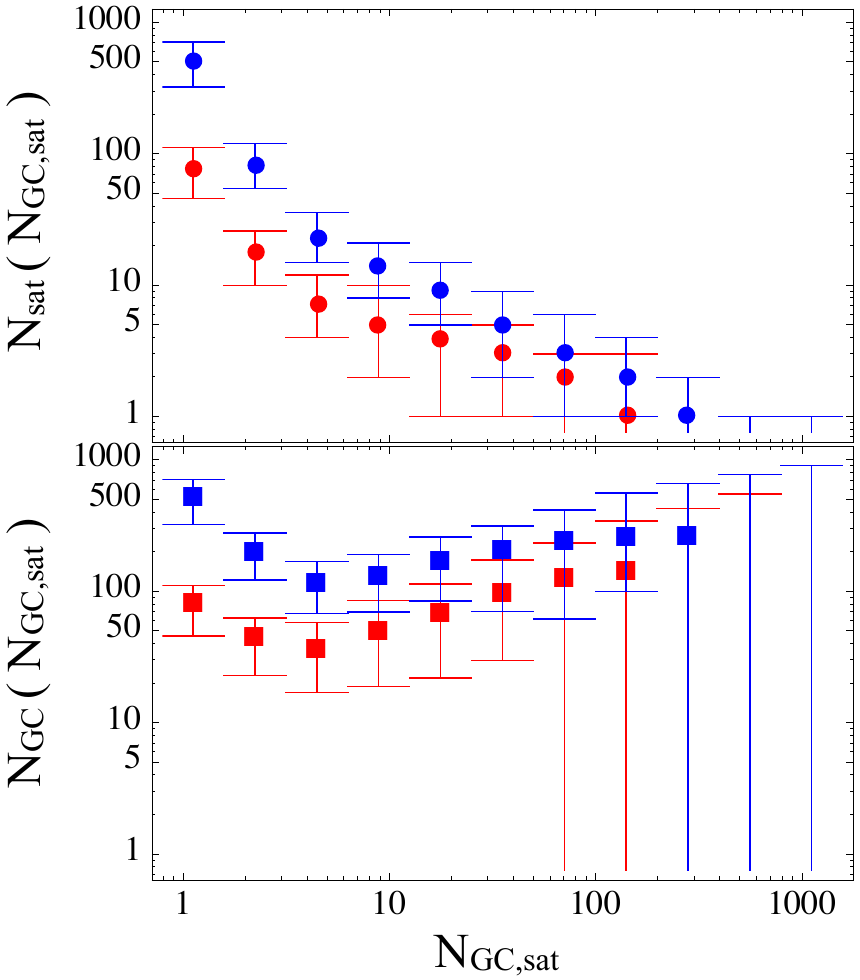}
\caption{{Upper panel: the total number of accreted satellites ($N_{\rm sat}$, on the $y$ axis) 
as a function of the number of GCs they contribute to the accreted population ($N_{GC, {\rm sat}}$, on the $x$ axis). Lower panel: the total number of accreted GCs ($N_{GC}$ on the $y$ axis) with progenitor satellites contributing $N_{GC, {\rm sat}}$ GCs (on the $x$ axis) to the accreted population.} Red and blue points refer to the populations of red and blue GCs. Points indicate median values over a sample of $10^3$ halo assembly histories, error bars extend between the 10 and 90\% quantiles.}
\label{fig1bis}
\end{figure}
\begin{figure*}
\centering
\includegraphics[width=.9\textwidth]{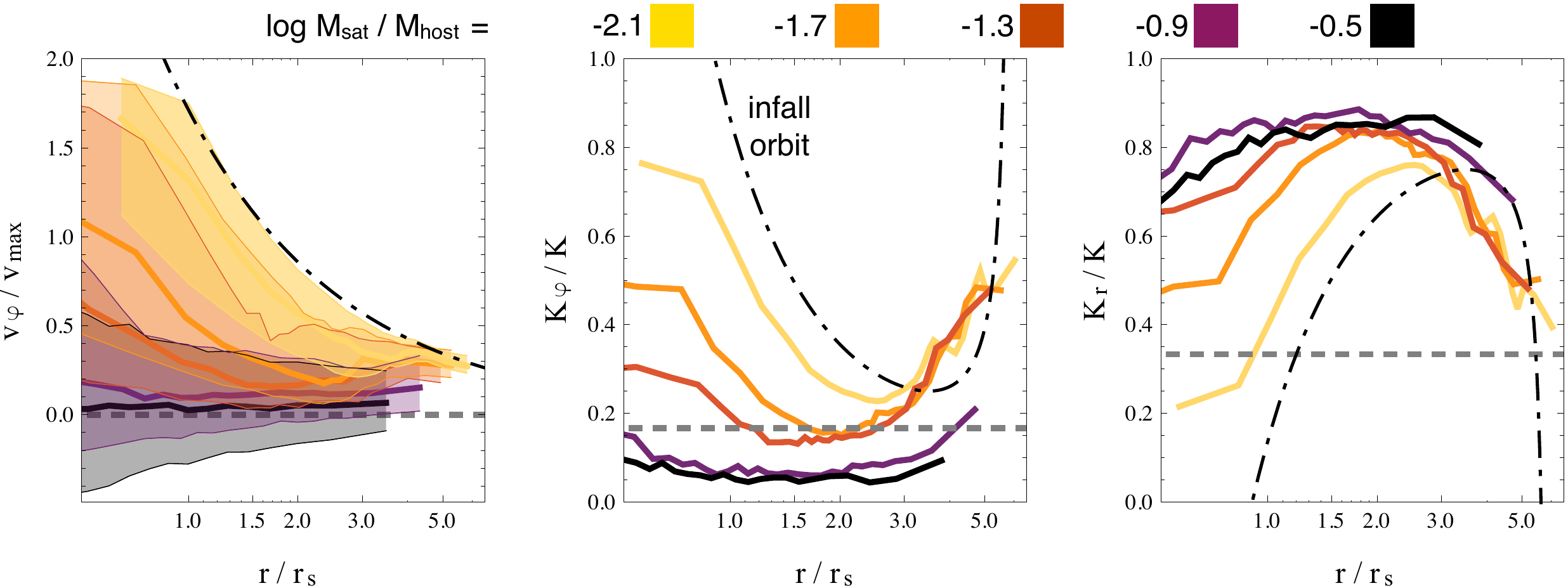}
\caption{Kinematical properties of individual contributions from satellites with different virial mass ratio at accretion, 
color-coded as in the legend at the top. The shaded regions in the {left} panel extend between 
the 10 and 90\% quantiles of the distribution 
of rotational velocities of particles contributed by individual satellites. The middle panel shows the fraction of the 
total kinetic energy in ordered rotation $K_\varphi /K$, the right panel the fraction in radial motion $K_r /K$. In all 
panels the dot-dashed black line describes the common infall orbit of all satellites, while the grey dashed line 
displays the expectation for a non rotating, isotropic population (after averaging over radius).  }
\label{fig2}
\end{figure*}

\section{Virial mass ratios and kinematics}

\citet{NA17b} (hereafter NA17) has shown that, for $\Lambda$CDM accretion
events, the satellite-to-host VMR is the most important single parameter to drive the kinematical properties
of the material accreted in minor mergers. The efficiency of dynamical friction increases with VMR, so that
massive satellites manage to sink deeper in the central regions of the host, and deposit
their contributions closer to the center. Instead, dynamical friction has little effect on the orbits of low-mass satellites,
especially for VMRs $\mr\lesssim1/50$.
As a result, a clear radial gradient emerges such that the VMR of the mean progenitor decreases with radius: i.e. minor 
mergers contribute preferentially at large galactocentric distances \citep[see also][]{RG16}.

NA17 also shows that massive satellites have their orbits radialised by dynamical friction, to the point 
that almost all memory of the orbital properties at infall is lost for VMRs $\mr\gtrsim1/20$. 
Consequently, the material deposited by massive satellites has a strong radial bias. 
On the other hand, minor mergers are substantially less affected by radialization, and preserve
a larger fraction of their orbital angular momentum. The mean orbital circularity of cosmological 
accretions is $j\sim0.5$ \citep[e.g.,][here $j$ is the ratio of the orbital angular momentum $J$
to the maximum angular momentum for the same orbital energy, $J_{\rm circ}(E)$]{AB05,AW11,LJ15}. Therefore,
on average, the material contributed by low mass satellites has a considerable 
angular velocity (see e.g. Fig.~10 in NA17).

We take the isolated minor merger N-body simulations presented in NA17. These consider the  
mergers of two idealised spherical, non-rotating dark matter haloes with Navarro-Frenk-White 
density profiles \citep[NFW,][]{JN97}. The suite presented in NA17 considers a range of different 
structural and orbital parameters, and we refer the reader to that work for more details. For this paper, 
we use the runs corresponding to mean values for the satellite-to-host density contrast (i.e. mean 
dark halo concentration parameters for both host and satellite haloes), a fixed median circularity at infall of $j=0.5$ and
a fixed orbital energy at infall.
We consider a range of values for the satellite-to-host VMR: $\mr\in\{-2.1,-1.7,-1.3,-0.9,-0.5\}$.
As discussed in NA17, the limits of this approach are analogous to those of the so called `particle-tagging'
technique \citep[e.g.,][]{JB01,NN03,BJ05,AC10,AC13,RA17}, which are extensively discussed in the 
literature \citep[see e.g.,][NA17]{AC10,JBa14,AC17}. Here we use a tagging fraction of $f_{\rm tag}=8\%$.

The left panel of Fig.~3 shows the distribution of orbital velocities $v_\varphi$ of material 
contributed by satellites with different VMR. The angle $\varphi$ is the angle in the satellite's 
orbital plane, $r_{\rm s}$ is the scale radius of the host NFW halo and $v_{\rm max}$ is its 
maximum circular velocity. The shaded regions extend between the 10 and 90\% quantiles 
of the distribution and the colour-coding is as in the legend at the top. 
The kinematical properties of the accreted material display a clear gradient with VMR.
Material contributed by massive satellites does not preserve coherent rotation, and is scattered 
in an approximately symmetric way with respect to the direction of the satellite's orbital motion at infall, i.e. with respect to $v_\varphi=0$. 
With decreasing VMR, more and more angular momentum is retained and the 
contributed material traces more closely the infall orbit itself, which is shown as a dash-dotted line. 

The middle and right panels in the same Figure show the fraction of the total kinetic energy 
$K$ associated with rotating motion in the $\varphi$ direction,
$K_\varphi/K$, together with the fraction in radial motion $K_r/K$. Lines display medians over the 
contributed material. Horizontal dashed lines show the expectation for the mean fraction of kinetic energy of a non rotating 
population with isotropic orbital distribution. The dash-dotted lines indicate the infall orbit.
As in the case of the distribution in rotational velocity $v_\varphi$, the kinematical properties of
material contributed by lower and lower VMR becomes more similar to the orbit of the progenitor satellite.
With increasing VMR tangential motion decreases systematically due to dynamical friction, with a 
corresponding increase in radial bias.

\begin{figure}
\centering
\includegraphics[width=\columnwidth]{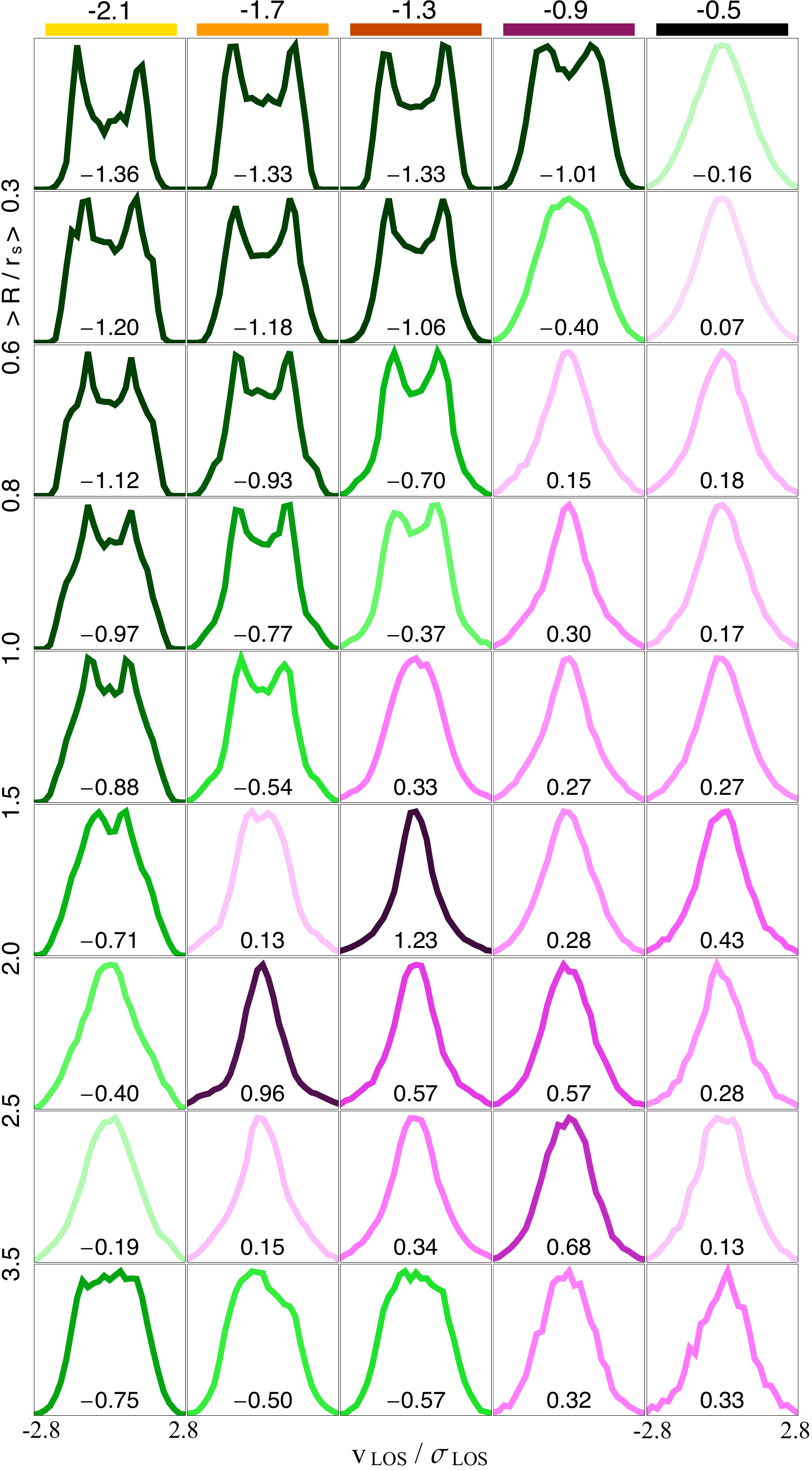}
\caption{Line-of-sight velocity distributions for accreted populations contributed by satellites with 
identical satellite-to-host virial mass ratio. Each column refers to the virial mass ratio shown at the 
top. Different rows refer to different projected radial intervals. Individual line profiles are color-coded 
by the value of the kurtosis $\kappa$, which is indicated at the bottom of each panel. }
\label{fig3}
\end{figure}

\subsection{LOSVDs and kurtosis}

The kinematic properties illustrated in the previous Section characterize the material
deposited by individual satellites. Here we wish to focus on the LOSVDs obtained by 
superposing the contributions of multiple satellites.
For the moment we ignore the cosmological setting (which will be introduced in Section~4) and 
for each VMR we superpose a large number of identical contributions, from identical satellites infalling 
on identical orbits with random directions. Figure~4 shows the resulting LOSVDs: different columns
identify different mass ratios as indicated at the top, while different rows are for 
different projected circular annuli, in units of the characteristi radius of the host NFW profile, $r_s$. 
All LOSVDs are displayed in terms of the normalised LOS velocity 
$v_{\rm LOS}/\sigma_{\rm LOS}$, where $\sigma_{\rm LOS}$ is calculated in the same radial interval.
Profiles are colour-coded according to the value of their kurtosis $\kappa$, which is also
displayed at the bottom of each panel. Green colours identify negative values of $\kappa$, 
while LOSVDs coloured in shades of pink have $\kappa>0$. A diagonal colour divide is evident:
negative values of the kurtosis are common for low VMRs and closer to the centre of the host.
Positive values of the kurtosis become prevalent for increasing VMRs and at larger radii. 
For context, assuming mean concentration parameters \citep[e.g.,][]{AL14}, 
an NFW halo with $\log\Mh(z=0)/\Mo=13.5$ has a scale radius $r_s$ of $r_{\rm s}\sim70~$kpc ($r_{\rm s}\sim100~$kpc) at $z=1$ ($z=0$). {Note that the radial range covered here is
substantially larger than in typical observations.}

This qualitatively shows that negative values of $\kappa$ have their origin in the contribution of minor mergers.
In particular, markedly double peaked LOSVDs are common in the upper-left corner. As showed in the previous Section, material contributed by 
minor mergers follows quite closely the progenitor's orbit, with high angular velocity. Double peaks are
then caused by the superposition of individual sub-populations rotating with random orientations. 
With increasing VMR, the degree of tangential motion in the contribution of each single satellites
decreases, causing the double peaks to disappear. Higher virial mass ratios result in LOSVDs with heavy tails
and sharper central peaks, as expected for material moving on radially biased orbits. 

Across different VMRs, the kurtosis $\kappa$ have a similar trend with radius. i)
$\kappa(R)$ is increasing at small radii; ii) reaches a maximum that depends on VMR at a radius that decreases with VMR;
iii) $\kappa(R)$ appears to decrease at large radii. 
This behaviour tracks orbital structure. At small (large) radii most contributed particles approach 
pericenter (apocenter), making double peaks more pronounced and lowering the value of the kurtosis. 
Due to dynamical friction, the radius of these turnaround points increases with decreasing VMR.
Finally, as massive satellites shed material that fills much larger phase space volumes, 
radial gradients in $\kappa(R)$ are less pronounced for high the highest VMRs.

In Fig.~4 we have superposed the contributions of a large number of identical satellites.
While illustrative, this choice makes the resulting LOSVDs unrealistic. The associated values of  $\kappa$ are somewhat extreme
when compared to values usually achieved in the context of distribution function based dynamical modelling \citep[e.g.,][]{OG91,OG93,vdM93,MC95,GB97}.
For example, very sharp double peaks are apparent, due to artificially identical distributions of pericenters and apocenters.
This effect is diluted in a cosmological context, where satellites accreted at different redshift deposit material with different orbital energies and at different radii around the host
\citep[e.g.,][NA17]{BJ05,AC10}.

There is however a second genuine difference between the phase space distribution of 
halo populations contributed by minor mergers and the phase space distributions generally used 
for modelling the main stellar body of galaxies. The latter are usually required to generate 
density distributions that are monotonic with galactocentric distance, corresponding to monotonic
distributions in energy. When seen in projection, these are characterized by a superior abundance of material 
moving with low LOS-velocity, and therefore more clearly bell-shaped LOSVDs. 
In turn, each minor merger contribution represents a coherent phase space pocket, with a tight and non-monotonic 
distribution in energy. Depending on the value of the VMR, this corresponds to a dearth of material with low energies
and LOS-velocity, as well as to central density hole (see NA17).

\section{The kinematics of cosmologically accreted GC populations}

We now combine the $\Lambda$CDM assembly histories presented
in Section~2 (a subset of 200) to the minor merger simulations introduced 
in Section~3, to construct toy models of the GC populations cosmologically 
accreted by galaxies hosted by a halo with $\log \Mh(z=0)/\Mo=13.5$.
As done in Section~2, we track blue and red GCs accreted since $z=6$ separately, and
assume that accreted satellites have deposited all of their GCs in the accreted halo of the host.  
We use N-body simulations with different VMRs to assign phase space coordinates to each GC at $z=0$, 
sampling randomly among the most bound satellite's particles, up to a tagging fraction of $8\%$. 
We assume accretion events are uncorrelated, and that satellites
infall from random directions. To account for redshift evolution, we scale $r_{\rm s}$ 
and $v_{\rm max}$ to the values of the host NFW halo at the accretion redshift, 
using our assembly histories and a mean mass-concentration-redshift 
relation \citep[e.g.,][]{LG08,AL14}. We do not expect such a simplified model 
to provide a realistic description of the accreted GC populations.
Rather, this investigation is aimed to explore to what degree the combination of the fundamental 
ingredients of the process of cosmological assembly 
(i.e. $\Lambda$CDM assembly histories, gravitational dynamics
and observed GC abundances) can reproduce the observed kinematic properties and trends
of GC halo populations.

\begin{figure*}
\centering
\includegraphics[width=\textwidth]{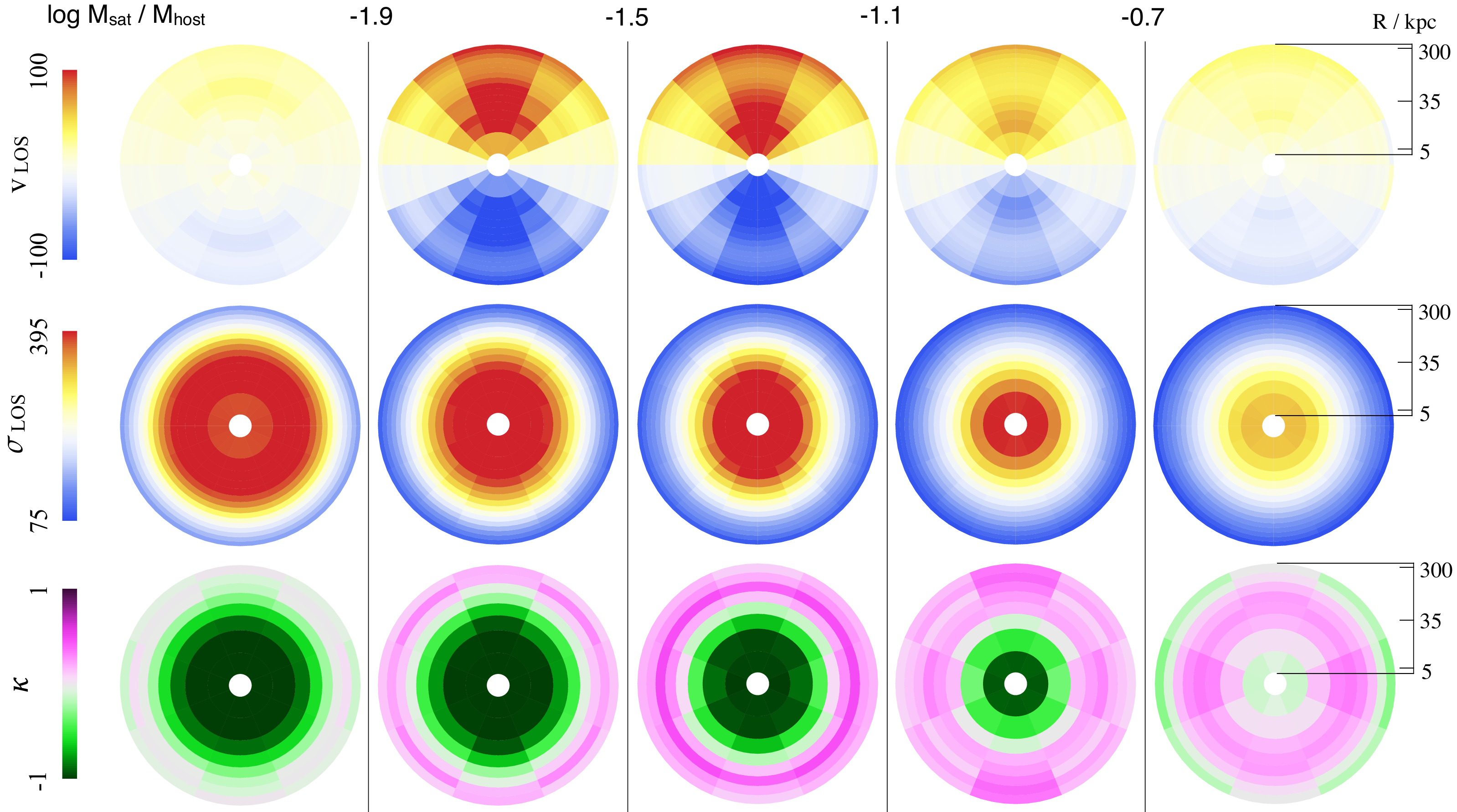}
\caption{Edge-on projected kinematic maps for the accreted blue GC population (medians 
over 200 accretion histories), for a massive galaxy with $\log M_{host}(z=0)/M_\odot=13.5$. 
Contributions of different satellites are grouped according to the value of the virial mass ratio at accretion.
{The legend at the top of the figure indicates the limiting values that define the considered 
bins in virial mass ratio}. }
\label{fig5}
\end{figure*}

\subsection{Edge-on kinematic maps by virial mass ratio}

Figure~5 shows maps of the LOS kinematics of cosmologically accreted blue GCs. 
Different columns refer to sub-populations of GCs contributed since z=6 by satellites with different 
VMRs, with values as in the legend at the top. {The corresponding maps for the red 
GCs would be essentially indistinguishable due to the small size of the adopted VMR bins}\footnote{{
Note that the full populations of blue and red GCs will still be different from each other due to the different relative 
contributions of the considered bins.}}. The maps show median values across a set of 200 individual assembly histories.
The top row shows 
values of the LOS velocity $v_{\rm LOS}$, the middle row shows velocity dispersion 
$\sigma_{\rm LOS}$, the bottom row displays values of the kurtosis $\kappa$. Note that
the radial scale is logarithmic, extending between 3 and 350~kpc as illustrated in the right-most panels.
To preserve any rotation pattern, the displayed maps are edge-on projections. For each individual 
assembly history and bin in VMR, a edge-on projection is obtained by aligning with respect to the 
corresponding total angular momentum. {In addition, symmetry with respect to the axis of rotation 
is assumed. }

First, we notice that the kurtosis $\kappa$ displays very similar patterns as those identified 
in Section~3.1. i) Negative values of $\kappa$ are dominant for material contributed by 
minor mergers. ii) $\kappa$ increases with VMR and is higher at larger galactocentric radii. 
Both of these trends survive the `mixing' process that accompany the cosmological setting,
which causes material accreted at higher redshift to be deposited closer to the centre of the 
host halo.

Interestingly, the superposition of the contributions of multiple satellites may still result in coherent 
rotation, despite the assumption of uncorrelated infall directions. Negligible residual rotation 
is seen in the GC sub-population contributed by the satellites with the lowest or highest VMRs.
In the former, the numerous individual contributions result in close to complete averaging. 
In the latter, each satellite deposits a non-rotating contribution, as shown in Fig.~3. 
At intermediate VMRs, however, material deposited by each individual satellite retains 
substantial rotation and the number of uncorrelated contributions is low enough that coherent 
rotation may survive. The mean rotation patterns seen in the averages of Fig.~5 are therefore 
dominated by the driving signal associated with the contribution of one single satellite, or to the 
fortuitous constructive superposition of a very small number of them. The significant scatter 
in the relations~(\ref{NGCrel}) may also help individual contributions to dominate within their VMR bin. 
Note however, that any residual rotation in the two bins $-1.9<\log\mr\leq-1.5$ and $-1.5<\log\mr\leq-1.1$ 
will likely not add up in a constructive way, as the chance of aligning angular momentum vectors is low. 

Values of the LOS velocity dispersions are clearly decreasing with VMR, a consequence 
of the increasing efficiency of dynamical friction. Material sinking to the center looses energy and 
has therefore lower velocities (see also NA17). This results in radial segregation, as shown in 
Figure~6, displaying the cumulative distribution functions (cdfs) of the projected number counts of the 
different GC sub-populations. Thick lines are medians over the considered 200 assembly histories and
the shaded regions extend between the 10 and 90\% quantiles. The different sub-populations are radially 
ordered by VMR. Scatter around the median profiles increases with VMR, due to the increasing Poisson noise resulting 
from a smaller number of contributions for higher VMRs. 
Blue GCs accreted by satellites with $\log\mr>-0.7$ have a half-number radius $R_{\rm h}$ between 13 
and 43~kpc (10 and 90\% quantiles); between 135 and 185~kpc for $\log\mr<-1.9$. The corresponding values for red GCs are very similar: the sub-populations of red and blue GCs defined by VMR of the 
contributing satellites are essentially indistinguishable. It is their different fractional contribution that 
causes any differences in the global kinematic properties of blue and red GCs, which we analyze in the following.

\begin{figure}
\centering
\includegraphics[width=.8\columnwidth]{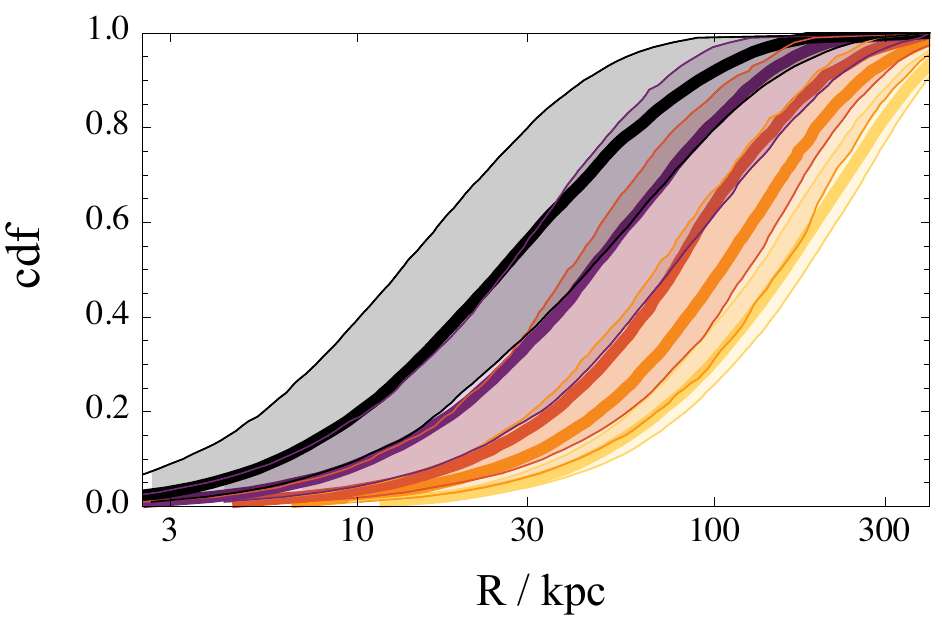}
\caption{The cumulative distribution for the projected number counts
of the sub-populations of cosmologically accreted GCs contributed by satellites with different mass ratios. 
Color-coding as in Fig.~4. The shading extends between the 10 and 90\% quantiles over the studied
set of 200 assembly histories. }
\label{fig4}
\end{figure}
\begin{figure*}
\centering
\includegraphics[width=\textwidth]{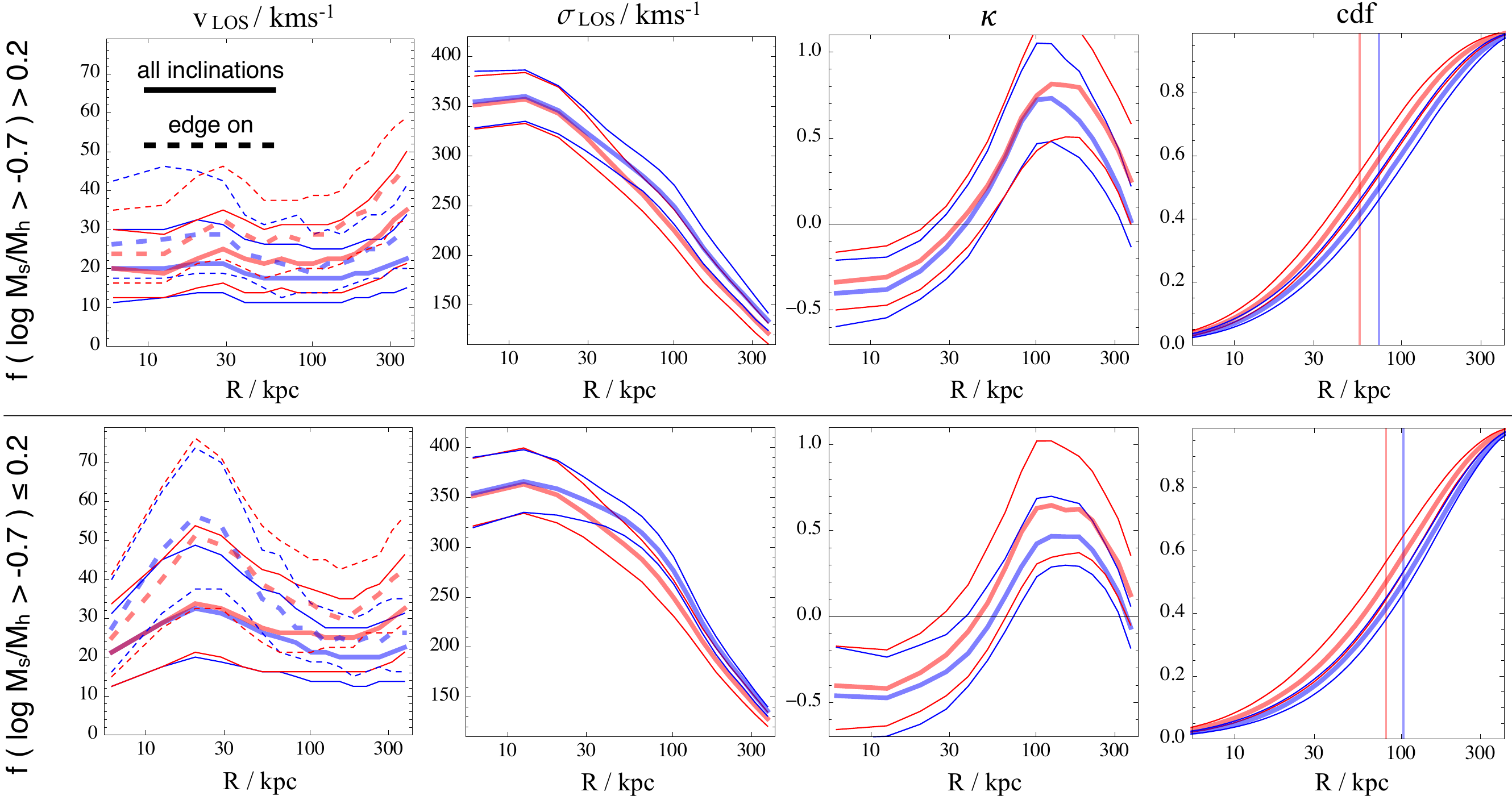}
\caption{Mean projected kinematics for the cosmologically accreted blue and red GC populations, 
as a sum over all contributing satellites. The two rows of panels display the effect of 
different accretion histories: upper (lower) panels refer to assembly histories in which
accretion events with $\log\mr>-0.7$ contribute $>20\%$ ($\leq20\%$) of all blue GCs,
corresponding to a 50\% of all studied assembly histories. In each panel, full lines 
average over viewing angles. Dashed lines in the left panel refer to viewing angles close
to edge-on with respect to the total angular momentum of the accreted blue GCs (see text for details). 
In each row, the first panel shows the maximum line-of-sight rotational velocity, 
the second panel displays line-of-sight velocity dispersion, the third panel shows kurtosis, 
the forth panel displays the cumulative distribution function for the projected 
number counts. In all cases, the sets of lines display the 25, 50 and 75\% 
quantiles of the distribution, over different accretion histories and inclinations. 
The vertical lines in the right-most panel identify the half-count radius.}
\label{fig6}
\end{figure*}

\subsection{Mean projected kinematics}

Figure~7 shows the LOS kinematic profiles of the full blue and red populations of 
accreted GCs. The sets of lines show the 25, 50 and 75\% quantiles over our set of 
200 assembly histories. Full lines include an averaging over inclinations (with respect to the 
angular momentum of the full blue GC accreted population), which we perform
by using 20 random viewing angles for each individual halo. From left to right, columns 
display respectively: maximum rotational velocity $v_{\rm LOS,max}$, velocity dispersion 
$\sigma_{\rm LOS}$, kurtosis $\kappa$ and cdf of the number counts. 

We mimic the measurement procedures used with real data \citep[e.g.,][]{VP13}: values of $v_{\rm LOS,max}$ 
are obtained by performing azimuthal fits in different radial bins, assuming a simple sinusoidal dependence,
\begin{equation}
v_{\rm LOS}(R,\theta)=v_{\rm LOS,max}(R)\cos(\theta-\theta_0) \ .
\label{rot}
\end{equation} 
To simulate the effects of differing LOS orientations, this procedure is repeated for 20 random viewing angles. 
Full lines display quantiles over the full set of angles, dashed lines simulate a close to edge-on projection, 
by showing quantiles over the subset of those 10 angles that result in the highest values of $v_{\rm LOS,max}$.

The two rows in Fig.~7 focus on a total of 100 different assembly histories each, dividing
the full set according to the fraction of GCs contributed by satellites with $\log\mr>-0.7$. 
As seen in Fig.~2, despite the importance of minor mergers, accretion events with high VMR are highly stochastic, and may contribute a substantial fraction of the 
accreted GC population. In 50\% of our assembly histories, the fraction of blue 
GCs contributed by satellites with $\log\mr>-0.7$ is $<0.21$: $f_{50}(\log\mr>-0.7)=21\%$.
This threshold separates the assembly histories contributing to the top/bottom panels of Fig.~7. 
Stochasticity, however, is such that, in a 5\% of all assembly histories, satellites with $\log\mr>-0.7$ contribute as much as 57\%
of all blue GCs: $f_{95}(\log\mr>-0.7)=57\%$. 

There are few aspects worthy of notice.
\begin{itemize}
\item{First of all, the accreted blue GCs are systematically more spatially extended than the red GCs, 
in both families of assembly histories. 
At the same time, the velocity dispersion of blue GCs stays consistently higher than that of red GCs.
Both of these aspects are in agreement with observations and, in these models, are a natural consequence of the slightly different
abundances prescribed by eqns~(\ref{NGCrel}). Blue GCs have a higher fractional contribution from minor mergers, and therefore
more extended density distributions and higher velocity dispersions.} 
\item{Slightly more in detail, over the 200 explored assembly histories, we find that 
the 10-to-90\% quantile range for the projected half-number radius of the blue GC population is 
$60-111$~kpc, while the red GCs have 45-94~kpc. These correspond respectively to 
intervals of $\approx9-17\%R_{vir}$ and $\approx7-14\%R_{vir}$, in good quantitative agreement 
with recent compilations by \citet{MH17} and \citet{DF17b}.}
\item{The median radial profile of the kurtosis $\kappa$ has a well defined shape, 
with preferentially negative values for both blue and red GCs within $\sim40$~kpc 
($\sim50$~kpc in the bottom row, when minor mergers are dominant). Then, $\kappa$ quickly increases towards 
positive values at larger radii. If the contribution of high VMR accretion 
events is substantial, higher values of $\kappa$ are achieved (top row).}
\item{Differences in the kinematical properties between the bottom and top row are systematic, but 
limited with respect to the scatters introduced by the details of the assembly histories. 
Residual rotation is stronger when minor mergers are dominant, and the velocity dispersion  
of both populations is consistently higher. }
\item{The most noticeable difference between the bottom and top rows
is in the spatial distributions. Vertical lines in the rightmost panels indicate half-count radii, which 
are significantly larger when the contribution of minor mergers is higher. 
We can not make a direct comparison between these predicted sizes and observations, as
these models do not account for the contribution of the {\it in situ} GC population. We notice however 
that the half-count radii displayed in Fig.~7 compare well with observations of galaxies with similar halo mass \citep[e.g.,][]{DF17b,MH17}.
The correlation between size and assembly history is also especially promising.}
\end{itemize} 

Finally, we note that rotation patterns in the GC populations of individual haloes are often twisted, i.e. 
the position angle $\theta_0(R)$ as defined in eqn.~(\ref{rot}) varies significantly with radius. This is 
the case for both blue and red GC populations. At the same time, red and blue GCs often have 
different kinematic position angles. This is a consequence of the stochastic nature of residual 
rotation in these simple models \citep[see also e.g.,][]{Moo14}. As mentioned in Section~4.1, 
coherent rotation is driven by the contributions of satellites with intermediate VMR and survives 
only in the absence of efficient averaging. Due to both intrinsic and Poisson noise on the 
relations~(\ref{NGCrel}), it is not unlikely that the rotation signal of blue and red GC populations 
is dominated by the contribution of different satellites, with likely misaligned angular momentum vectors.

\begin{figure}
\centering
\includegraphics[width=.85\columnwidth]{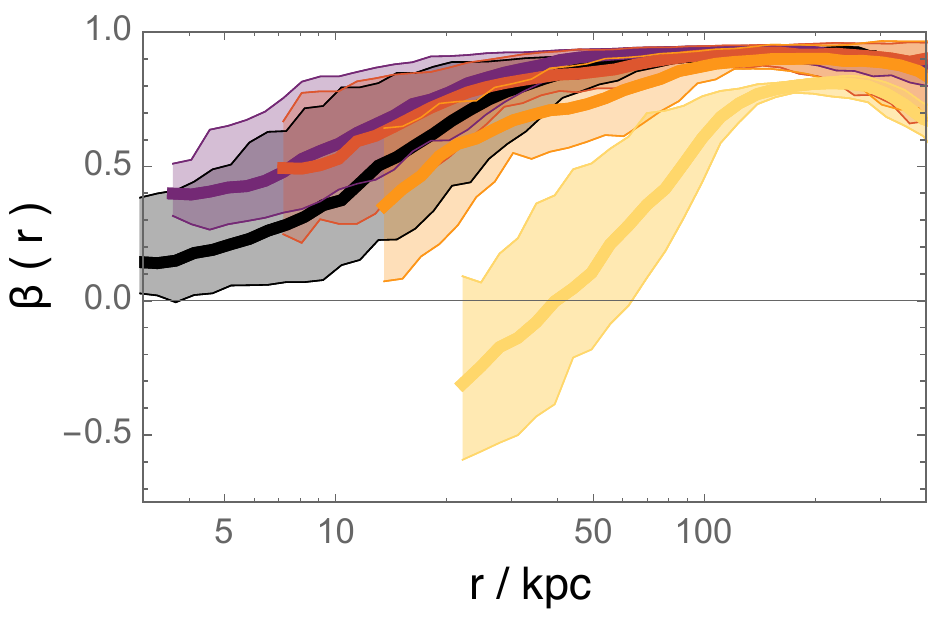}
\caption{Profiles of the anisotropy parameter $\beta(r)$ for the GC sub-populations as in Fig.~5 and~6.
Lines indicate median values and the shading extends between the 10 and 90\% quantiles over the studied
set of 200 assembly histories. Colour-coding as in Fig.~4 and~6.}
\label{fig4}
\end{figure}

\subsection{Orbital anisotropy}

We have shown that negative values of the kurtosis are a natural product of the residual 
tangential motion of the contributions of minor mergers and that they are ubiquitous in the 
central regions for the accreted population of GCs. Here, we 
address orbital structure. Figure~8 shows the anisotropy profile $\beta(r)$ of the 
sub-populations of the accreted blue GCs defined by VMR of the parent satellite, as in 
Figs.~5 and~6, with the same colour-coding. Thick lines show medians over our 200 
assembly histories, shaded regions extend between the 5 and 95\% quantiles of the distribution. 

For all VMRs, anisotropy profiles increase towards large radii, with very high values
of $\beta$ reached at $r\gtrsim100~$kpc. Due to the segregation induced by dynamical friction,
the profiles relevant to low VMRs do not extend to the very central regions of the host, 
where not enough GCs are contributed. Except for the contribution of satellites with $\log\mr<-1.9$,
the other subpopulations display a positive $\beta$ over the entire radial range.
In the very central regions, the anisotropy of the sub-population contributed by the highest VMR satellites
is mildly radial, $\beta(<20~{\rm kpc})\sim0.3$. This value increases for intermediate VMRs, but 
quickly decreases to negative values of $\beta$ for the subpopulation with the lowest VMR. 
Although the scatter is considerable, we find that 
the superposition of GCs cosmologically accreted from satellites with $\log\mr<-1.9$
is approximately isotropic at $r\sim40~$kpc and {\it tangentially biased} at smaller radii. 
Values of the anisotropy that are as low as $\beta\sim-0.5$ are within the 5\% quantile.
This tangential bias, however, is shared by only a very small fraction of the GCs 
in this sub-population: as mentioned in Sect.~4.1, the half-count 
radius of the GCs accreted from satellites with $\log\mr<-1.9$ is $R_{\rm h}>135$~kpc in 90\% of cases. 
Only a fraction of 8\% is at $R<40$~kpc. While tangential anisotropies are not uncommon
in the very central regions for the sub-population contributed by minor mergers, 
this is not predicted to be a dominant behaviour.

\begin{figure}
\centering
\includegraphics[width=\columnwidth]{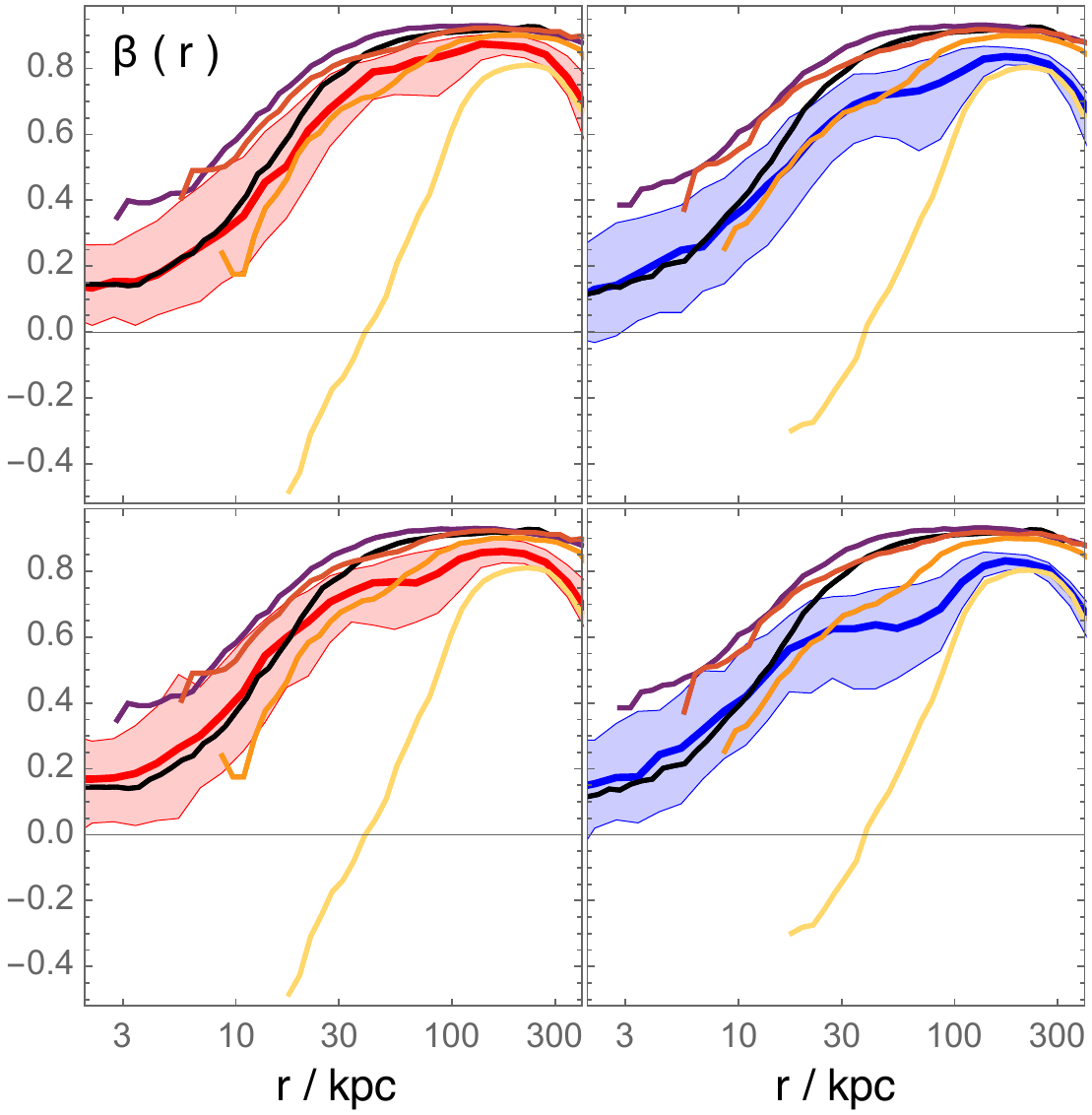}
\caption{The anisotropy profiles of the cosmologically accreted red (left panels) and blue
(right panels) GC populations. Full lines show medians, the shading extends between the 
5 and 95\% quantiles of the distribution. Top and bottom row are for different assembly 
histories, as in Fig.~7. For comparison, panels also display the anisotropy profiles of the
sub-populations defined by virial mass ratio, as in Fig.~8. }
\label{fig7}
\end{figure}

Figure~9 shows the anisotropy profiles of the full populations of red (left panels)
and blue GCs (right panels). Full lines show medians, the shading extends between the 
5 and 95\% quantiles of the distribution. As in Fig.~7, the top row collects the 100 assembly
histories in which the fractional contribution to the blue GC population of satellites with 
$\log\mr>-0.7$ is $<0.21$. The bottom row has the complementary set with $f(\log\mr>-0.7)\geq0.21$.
The median anisotropy of the GC sub-populations defined by VMR are also shown,
for comparison, with the same colour-coding as in Fig.~8. We find that both red and blue
GC populations have predominantly radial anisotropy, and that any dependence on the
assembly history is minor. For all practical purposes, accreted red an blue populations
have very similar anisotropy profiles: $\beta$ is very mildly positive in the central regions 
($r\lesssim10~$kpc), and then displays a sustained increase towards larger radii, 
with little scatter across different assembly histories.
The anisotropy profiles of the full red and blue population can be seen to follow the profile of
the subpopulation with the highest VMR in the centre, and of the population with the lowest
VMR in the outermost regions, where they respectively dominate. Not enough GCs contributed
by satellites with $\log\mr<-1.9$ appear to be present in the very central regions to 
result in a tangentially biased orbital structure.

\begin{figure*}
\centering
\includegraphics[width=.9\textwidth]{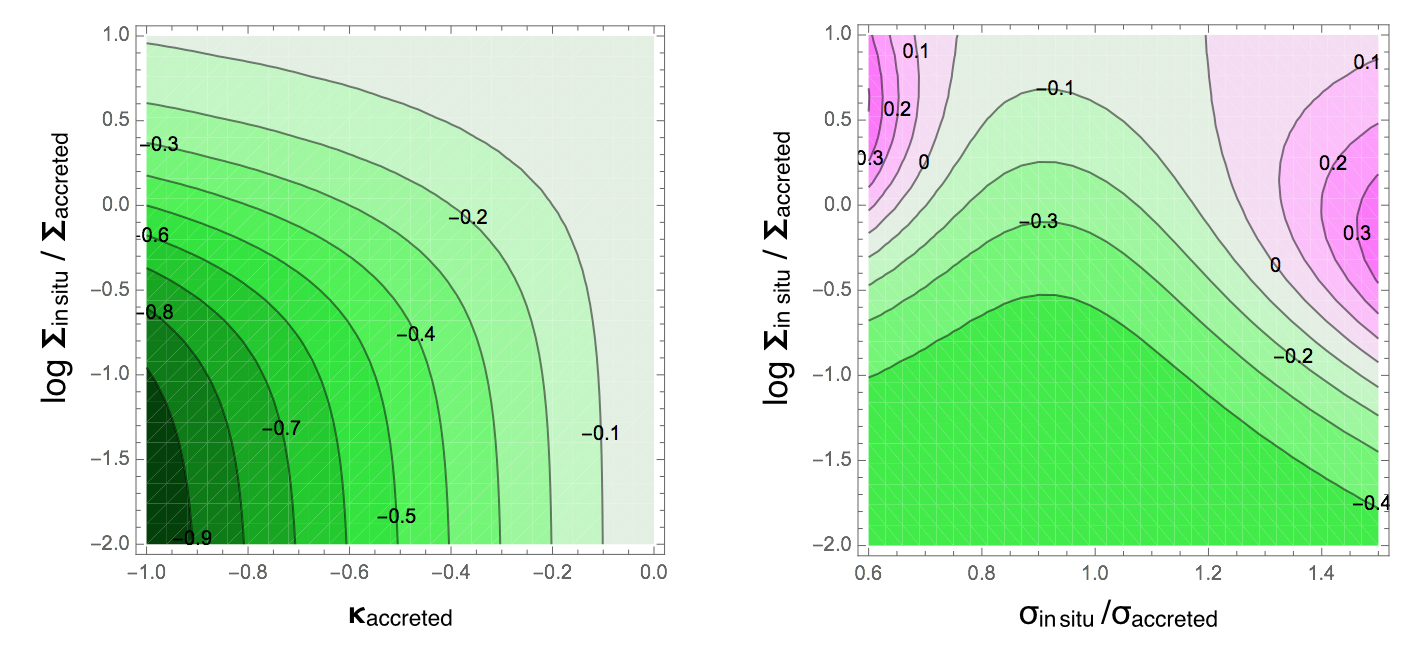}
\caption{The kurtosis $\kappa$ of a population including an in situ and an accreted component. 
(the LOSVD of the in situ component is assumed Gaussian). The left panel 
shows values of the kurtosis from the superposition of an accreted population with varying
$\kappa_{accreted}$ (on the x axis) and an in situ population with varying relative normalization (on the y axis). 
The LOS velocity distribution of the in situ component is kept fixed at $\sigma_{in situ}=\sigma_{accreted}$.
The right panel shows values of the kurtosis from the superposition of an accreted population with
$\kappa_{accreted}=-0.5$ and an in situ population with varying $\sigma_{in situ}$ and 
relative normalization (respectively, on the x and y axis).  }
\label{fig8}
\end{figure*}

\section{Superposing accreted and `in-situ'}

{
So far, we have dealt with the population of accreted GCs. As we have seen, once a $\Lambda$CDM 
framework has been chosen, their spatial distribution and kinematic properties can be constructed quite easily.
There is a very limited number of free parameters to fix: by concentrating on the accreted population alone, 
the current analysis remains a self-contained one. However, a direct comparison with 
observations would of course require a model for the population of `in-situ' GCs.
Unfortunately, it is not currently straightforward to derive the kinematic properties of the in-situ population in a constructive
manner, as done for the accreted population, as a model for its formation mechanism(s) is not readily available, and
in fact currently the subject of intense study \citep[see e.g.,][and references therein]{DF18b}.
We can certainly assume that the in-situ population is negligible in the outer regions, but to explore how the two components 
combine in the total observed GC population requires a full characterisation of both. }

{
In the absence of a constructive model for the in-situ component a number of arbitrary 
choices become necessary. Even under the most extreme simplification, we would need to explicitly fix:
 i) the total abundance of GCs formed in-situ (the fraction with respect to the 
abundances prescribed by eqns~3); ii) its half-number radius and detailed density profile;
iii) its full orbital structure, which includes its anisotropy as well as whether it has any ordered rotation. 
A specific formation mechanism would provide a prediction for the quantities above, an alternative possibility 
is to explore a range of choices for these. Such a detailed exploration goes beyond the scope
of the current study, but we can still gain substantial insight using a simplified framework. What we are most interested in here 
is in how the properties of the LOS velocity distribution of the accreted population get `diluted' by those of the in-situ population. 
In particular, in whether the negative values of $\kappa$ survive the superposition of an in-situ population.}

{We consider the observed LOSVD, $f_{LOS,tot}(v,R)$, i.e. the superposition}
\begin{eqnarray}
\Sigma_{tot}(R) f_{LOS,tot}(v,R) = & \Sigma_{ac}(R) f_{LOS,ac}(v,R) + \nonumber \\ 
& +\ \Sigma_{in}(R) f_{LOS, in}(v,R)\ .
\end{eqnarray} 
{Here we have taken that all LOSVDs $f_{LOS,j}(v,R)$ (for all $j$) are normalised to unity, $\int f_{LOS,j}(v,R) dv=1$, 
and that $\Sigma_{j}(R)$ is the surface density of the different components. We have seen that negative
values of $\kappa_{ac}$ are the norm in the inner regions ($R\gtrsim40$~kpc) for the accreted population. 
On the other hand it is plausible to assume that the LOSVD of the in situ population is close to normal, with 
$\kappa_{in}\approx 0$. As a simple representative case we take that $f_{LOS,in}(v,R)$ is exactly Gaussian, with zero mean. 
With this hypothesis, Figure~10 explores the kurtosis $\kappa_{tot}$ of the observed LOSVD $f_{LOS,tot}$
obtained by superposing an accreted population and an in-situ population with different properties. }

{The left panel explores the case of different values of the kurtosis $\kappa_{ac}$ of the accreted population (on the $x$ axis)
and of a varying relative contribution of the in-situ component, $\log \Sigma_{in}/\Sigma_{ac}$ (on the $y$ axis). 
We have assumed that $\sigma_{ac}=\sigma_{in}$ and, for the LOSVD of the accreted population we have used 
the family of symmetric, positive definite, LOSVDs constructed in \citet{NA12}. Where the 
accreted component dominates, $\Sigma_{in}\ll\Sigma_{ac}$, $\kappa_{tot}\approx\kappa_{ac}$, as can be expected in the outer regions of the halo. 
The negative values of $\kappa_{ac}$ get progressively diluted by an increasing in-situ component. 
However, the left panel of Fig.~10 shows that $\kappa_{tot}$ remains negative even when the contribution of
the in-situ component is comparable to the one of the accreted component, and in fact even when higher.}

{Finally, in the right panel we relax the hypothesis that $\sigma_{ac}=\sigma_{in}$: while it is likely
that $\sigma_{ac}\approx\sigma_{in}$, the two values will in general be different, as the two components
satisfy different equilibrium conditions. As a representative value, we fix $\kappa_{ac}=-0.5$, which we 
have found to be typical for the innermost regions of the accreted population (see Fig.~7). Once again, 
we see that the contribution of an in-situ population acts to dilute the negative kurtosis of the accreted population. 
However, in most of the parameter space $\kappa_{tot}\gtrsim 0$. }

{
Instead of restricting ourselves to specific kinematical models, here we have explored the kurtosis resulting
from the superposition of two different LOSVDs, covering a region of parameter space that plausibly encompasses a variety of useful cases. 
A more detailed analysis would be useful to better elucidate the interplay between the kinematic properties of in-situ and accreted population,
in particular when these are required to both be in equilibrium. 
On the other hand, the exercise above already shows that, in the presence of a significant fraction of accreted GC population, 
negative values of the kurtosis are especially resilient. Fig.~10 demonstrates that, where $\kappa_{ac}\lesssim 0$, it is highly likely 
that the observed kurtosis $\kappa_{tot}$ is also negative if $\Sigma_{in}\lesssim\Sigma_{ac}$. The fact that negative values 
of the kurtosis are indeed observed for $\kappa_{tot}$, compatibly with our predictions for $\kappa_{ac}$, would appear to suggest
that the accreted GC population represents a sizable fraction of the total also at intermediate radii, $10\lesssim R/{\rm kpc}\lesssim40$.
This is also compatible with the fact that our accreted-only GC abundances do not fall short of the observed values (Sect.~2), and that the predicted half-number radii compare well with observations (Sect.~4.2). }

\section{Discussion and Conclusions}

In this paper we have investigated the kinematical properties of the GC populations 
assembled through hierarchical merging around massive galaxies. 
Our set up is purposely simplistic, and only accounts for the following fundamental 
ingredients: $\Lambda$CDM assembly histories, gravitational dynamics
and measured relations between GC abundances and dark matter halo mass. 
This is instrumental to highlighting the possible need for any additional physical 
processes that are not accounted for in this work. 

The ingredients above imply that the halo populations of accreted GCs have systematically 
different progenitors with respect to the stellar halo. This is due to the very different 
proportions with which accreted satellites contribute stars and GCs per unit of contributed dark matter. 
Because of the steepness of the SHMR, the stellar halo is dominated by the contributions of satellites 
with high VMR, which are strongly affected by dynamical friction. In contrast, if GC abundances are
approximately proportional to halo mass, minor mergers are substantially more 
important in building up the GC halo populations. 
Since the material deposited by minor mergers has different kinematic properties from 
the one contributed by satellites with high VMR, the accreted GC populations do not
in general trace the stellar halo. However, in hosts with $\log \Mh(z=0)/\Mo=13.5$, approximately $65\%$ 
of material in the stellar halo and in the accreted population of red GCs have been deposited 
by satellites with similar distributions of VMRs. This makes red GCs a better tracer of the stellar
halo. In turn, the contributions to the accreted population of blue GCs are systematically 
biased towards satellites with lower VMRs. In 75\% of $\Lambda$CDM assembly histories, the fraction of 
the accreted population of blue GCs contributed by satellites with $\mr<1/50$ at accretion is 
$>39\%$ ($>30\%$) for galaxies hosted by haloes with $\log \Mh(z=0)/\Mo=13.5$ ($\log \Mh(z=0)/\Mo=12.3$). 

With respect to the accreted stellar halo, the fingerprints of the increased importance of minor mergers for
the halo GC populations are: 
i) a more extended spatial distribution and ii) an increased degree of tangential motion. 
Even between red and blue GCs, differences in the relation between halo mass and GC abundance are 
sufficient to make the spatial distribution of blue GCs systematically more extended than the one of red GCs
(half-number projected radius in the interval 9-17\% $R_{vir}$ for the blue GCs and 7-14\% $R_{vir}$ in 80\% of cases), as well as to make the velocity dispersion of the former systematically higher. The increased degree of 
tangential motion makes LOSVDs with negative kurtosis a distinctive trait of the increased contribution 
of satellites with low VMRs, in no contradiction with the accretion picture.

The kurtosis profiles of our toy-model 
accreted GC populations are negative in the central regions $\kappa(R<40~{\rm kpc})\lesssim0$, 
and turn to positive values at larger radii, for both blue and red GCs. The galaxy-to-galaxy scatter is significant anf
details of each system depend on the specific assembly history. {In those regions
where the accreted population has $\kappa\lesssim 0$ the observed kurtosis can be expected 
to preserve a negative sign as long as the in-situ contribution is not largely dominant.} This 
justifies the fact that our predicted kurtosis profiles agree quite well with those observed \citep[e.g.,][]{VP13}.

These negative values of the kurtosis, however, do not correspond to tangentially biased orbital
distributions. The cosmologically accreted populations of GCs constructed using our set of simplifying hypotheses
have anisotropy profiles that increase from mildly radial in the central regions ($\beta(r<10~{\rm kpc})\sim0.2$) 
to strongly radially anisotropic at large radii ($\beta(r>30~{\rm kpc})\gtrsim0.6$). 
{This confirms that the correspondence between kurtosis and anisotropy is only qualitative.
A number of counterexamples to this correspondence are known in the literature \citep[e.g.,][]{OG93,vdM93,NN14},
here we have provided an additional example of an equilibrium population with $\kappa<0$, but $\beta\gtrsim0$ 
at some radius. }
We attribute this to the phase space properties of the material accreted from 
minor mergers, which is highly clumpy in the space of the integrals of the motion (see also NA17).
We find that the sub-population of cosmologically accreted GCs with VMR $\log\mr<-1.9$ 
is tangentially biased at radii $r\lesssim40~$kpc, where GCs are mainly approaching the 
pericenter of the cosmological infall orbits of their progenitor satellites.
Values of $\beta$ can be as low as $\sim-0.5$. However, at these radii, the contribution of
satellites with higher virial mass ratios is dominant in our models, and in fact drives the anisotropy 
of the full GC population, which remains positive in more than 90\% of cases.

\subsection{Comparison with dynamical analyses}

Measuring the orbital structure of pressure supported systems is a notoriously difficult task,
due to strong model degeneracies. The infamous degeneracy between mass and anisotropy  
makes it difficult to disentangle the orbital structure of the tracer population from the profile of the  
embedding gravitational potential, as well as from the details of density profile of the
tracers themselves \citep[see e.g.,][and references therein]{OG93,BT08,WE09,GM13}. Use of the 
higher moments of the LOSVD is helpful, though not always possible in dynamical analyses
of GC populations, due to the limited number of available tracers \citep[see e.g.,][]{NA12}.
As a consequence, literature results on the orbital structure of GC populations are mixed \citep[see e.g.,][]{AA14,NN14,Zhu14,Zhu16,ZP15,VP15,OA16,Wa17}. 
Additionally, a direct comparison with our results is not straightforward as many analyses have adopted 
models with radially constant orbital structure {(see however \citet{NN14} for an example of 
dynamical modelling with non-constant anisotropy)}. It is unclear whether a constant value of $\beta$ may 
correctly describe our model populations, and what value of $\beta$ would be inferred as a result. 
However, some studies have inferred a strong tangential orbital bias, which, despite these possible biases, 
appears difficult to reconcile with our simple models. 

There is a number of mechanisms that may possibly be responsible for this discrepancy. For instance, 
the following two physical processes, which we have not accounted for, may alter the kinematics of the 
GC populations in the central regions of the host.
\begin{itemize}
\item{First and foremost, we have neglected the contribution of GCs formed within the host halo itself, or formed
during mergers \citep[e.g.,][]{JK12} rather than simply accreted. These may be dominant in the 
central regions and our model can not predict their kinematics. {We have shown that 
an in-situ component in very unlikely to entirely erase the signal for a negative kurtosis, but we have not 
analysed how the in-situ component would affect the actual orbital structure and anisotropy of the 
full GC population.} The orbital distribution of material formed {\it in situ} is usually less radially biased 
than the one of material that is cosmologically accreted \citep[e.g.,][]{Ro14}. Additionally, these GCs may display disk kinematics, as seen in M31 \citep[e.g.,][]{NC16}, mimicking a tangentially biased orbital structure.}
\item{Second, we have ignored the possibility that some GCs may have been shredded by the tides after 
they are accreted onto the host. GCs that reach closer to the centre are more easily disrupted,
so that, as discussed in a number of works \citep[e.g.,][]{BW76,VP13,AA14,Wa17}, GC disruption 
may cause a systematic depletion of GCs on radial orbits, possibly altering the orbital distribution.}
\end{itemize}

It remains however unclear whether the latter effect may reconcile the present discrepancy. 
GC disruption is certainly a relevant process for low mass GCs \citep[e.g.,][]{GO97,Bau98,VH97,EV03,EP08,IG10,JK12,OG14,SM14}, but the accreted 
populations considered here only include those GCs that may survive in the contributing satellites.
Observations in the MW suggest that some GCs that likely have an accretion origin may indeed experience 
tidal disruption \citep[e.g.,][]{Od03,GD06}. On the other hand, our model does not clearly overestimate 
GC abundances, which would suggest that the fraction of GCs destined to disruption {\it after accretion}
is not substantial in our populations. This is supported by recent numerical simulations, which find that tides 
are in fact stronger in low mass hosts than are in massive galaxies \citep[][]{BK14,Zo16}.

In addition, both of the mechanisms above can only be effective in the central regions of the host:  
it appears difficult that they could significantly alter the strong radial bias displayed by our models at large radii. 
Therefore, observations of any signal for a tangential anisotropy at large galactocentric radii are especially interesting.
Despite the significant halo-to-halo scatter, our models are uniformly strongly radial at large radii. These 
observations therefore raise the question of whether a tangential anisotropy may be reproduced in a 
$\Lambda$CDM scenario by relaxing some of the hypotheses made here. For example, our models 
assume that most cosmological accretion events have intermediate circularity at infall ($j\sim0.5$), 
and that this value is approximately independent of mass ratio \citep[e.g.][]{AB05,AW11,LJ15}. 
As the material contributed by minor mergers best preserve memory of their infall orbit, 
the observed signal for tangential orbits may suggest differences in the orbital properties at infall. 
In future work, it would be interesting to explore whether a bias towards higher circularities for minor mergers
may in fact produce accreted GC populations that are less radially biased, or even tangentially biased at large radii.

To conclude, many fundamental properties of the kinematics of GC populations around massive galaxies appear to 
be simply interpreted as a direct consequence of the important role of minor mergers. 
In particular, we have shown that there is no contradiction between the accretion scenario and the 
observed negative values of the kurtosis, which are instead ubiquitous when accretion events with low VMRs are important. 
It will be interesting to compare the results of this work with those of future kinematical analyses.

\section*{Acknowledgements}
It is a pleasure to thank Jean Brodie, Duncan Forbes, Eric Peng and the Santa Cruz SAGES group for stimulating discussions.

\end{document}